\documentclass[aps,pra,amsmath,amssymb,superscriptaddress,reprint]{revtex4-1}

\usepackage{graphicx}
\usepackage{bm}
\usepackage{multirow}
\usepackage{verbatim}
\usepackage{longtable}
\usepackage{color}
\usepackage{multirow}
\usepackage{hyperref}
\usepackage[usenames,dvipsnames]{xcolor}
\usepackage[normalem]{ulem}

\usepackage{braket}

\begin{document}

\title{On-chip deterministic operation of quantum dots in dual-mode waveguides for a plug-and-play single-photon source}
\author{Ravitej Uppu}
\email{ravitej.uppu@nbi.ku.dk}
\affiliation{Center for Hybrid Quantum Networks (Hy-Q), Niels Bohr Institute, University of Copenhagen, Blegdamsvej 17, DK-2100 Copenhagen, Denmark}
\author{Hans T. Eriksen}
\affiliation{Center for Hybrid Quantum Networks (Hy-Q), Niels Bohr Institute, University of Copenhagen, Blegdamsvej 17, DK-2100 Copenhagen, Denmark}
\author{Henri Thyrrestrup}
\affiliation{Center for Hybrid Quantum Networks (Hy-Q), Niels Bohr Institute, University of Copenhagen, Blegdamsvej 17, DK-2100 Copenhagen, Denmark}
\author{Asl{\i} D. U\u{g}urlu}
\affiliation{Center for Hybrid Quantum Networks (Hy-Q), Niels Bohr Institute, University of Copenhagen, Blegdamsvej 17, DK-2100 Copenhagen, Denmark}
\author{Ying Wang}
\affiliation{Center for Hybrid Quantum Networks (Hy-Q), Niels Bohr Institute, University of Copenhagen, Blegdamsvej 17, DK-2100 Copenhagen, Denmark}
\author{Sven Scholz}
\affiliation{Lehrstuhl f{\"u}r Angewandte Festk{\"o}rperphysik, Ruhr-Universit{\"a}t Bochum, Universit{\"a}tsstrasse 150, D-44780 Bochum, Germany}
\author{Andreas D.~Wieck}
\affiliation{Lehrstuhl f{\"u}r Angewandte Festk{\"o}rperphysik, Ruhr-Universit{\"a}t Bochum, Universit{\"a}tsstrasse 150, D-44780 Bochum, Germany}
\author{Arne Ludwig}
\affiliation{Lehrstuhl f{\"u}r Angewandte Festk{\"o}rperphysik, Ruhr-Universit{\"a}t Bochum, Universit{\"a}tsstrasse 150, D-44780 Bochum, Germany}
\author{Matthias C. L\"{o}bl}
\affiliation{Department of Physics, University of Basel, Klingelbergstrasse 82, CH-4056 Basel, Switzerland}
\author{Richard J.~Warburton}
\affiliation{Department of Physics, University of Basel, Klingelbergstrasse 82, CH-4056 Basel, Switzerland}
\author{Peter Lodahl}
\affiliation{Center for Hybrid Quantum Networks (Hy-Q), Niels Bohr Institute, University of Copenhagen, Blegdamsvej 17, DK-2100 Copenhagen, Denmark}
\author{Leonardo Midolo}
\email{midolo@nbi.ku.dk}
\affiliation{Center for Hybrid Quantum Networks (Hy-Q), Niels Bohr Institute, University of Copenhagen, Blegdamsvej 17, DK-2100 Copenhagen, Denmark}

% ABSTRACT
\begin{abstract}
A deterministic source of coherent single photons is an enabling device of quantum-information processing for quantum simulators \cite{walther2012,nori2014,sciarrino2018}, quantum key distribution \cite{acin2018,zbinden2002}, quantum repeaters \cite{lodahl2019,lo2015}, and ultimately a full-fledged quantum internet \cite{kimble2008,hanson2018}.
Quantum dots (QDs) in nanophotonic structures have been employed as excellent sources of single photons \cite{Wang2016,Somaschi2016,Liu2018,Kirsanske2017,Wang2019} , and planar waveguides are well suited for scaling up to multiple photons and emitters exploring near-unity photon-emitter coupling \cite{Arcari2014} and advanced active on-chip functionalities \cite{papon_nanomechanical_2019}.
An ideal single-photon source requires suppressing noise and decoherence, which notably has been demonstrated in electrically-contacted heterostructures  \cite{Kuhlmann2015,Lobl2017,Thyrrestrup2018,Warburton2019}.
It remains a challenge to implement deterministic resonant excitation of the QD required for generating coherent single photons, since residual light from the excitation laser should be suppressed without compromising source efficiency and scalability.
Here, we present the design and realization of a novel planar nanophotonic device that enables deterministic pulsed resonant excitation of QDs through the waveguide.
Through nanostructure engineering, the excitation light and collected photons are guided in two orthogonal waveguide modes enabling deterministic operation. We demonstrate  a coherent single-photon source that simultaneously achieves high-purity ($g^{(2)}(0)$ = 0.020 $\pm$ 0.005), high-indistinguishability ($V$ = 96 $\pm$ 2 \%), and $>$80\% coupling efficiency into the waveguide.
The novel `plug-and-play' coherent single-photon source could be operated unmanned for several days and will find immediate applications, e.g., for constructing heralded multi-photon entanglement sources \cite{Rudolph2017} for photonic quantum computing or sensing.
\end{abstract}
\maketitle

The conventional approach to pulsed resonant excitation of a QD employs a cross-polarized excitation-collection scheme \cite{Somaschi2016,Wang2016,Liu2018}, which inherently limits the collection efficiency of the generated single photons to $\leq$ 50\%.
Recently, elliptical microcavities were proposed and tested to overcome this limit on efficiency \cite{Wang2019}, although this method is complicated by the need of controlling two narrow-band cavity resonances relative to the QD.
In comparison, planar nanophotonic waveguides offer broadband and robust operation and are naturally suited for efficient laser suppression since the excitation laser and the collection mode can be spatially separated, allowing to construct devices with near-unity generation efficiency.
However, resonant excitation of planar devices has so far relied on coupling the pump laser through leaky radiation mode \cite{javadi2018,Thyrrestrup2018,Grousson2014,Melet2008}, which results in high alignment sensitivity, uncontrolled specular scattering, and incompatibility with fiber coupling.
To overcome these limitations, the QD is ideally excited resonantly directly through the waveguide mode.
Here we demonstrate a tailored nanophotonic device that enables resonant excitation launched through a grating coupler into a waveguide and subsequent outcoupling of highly coherent single photons from the chip with an additional grating coupler.
The challenge for such an in-line approach is to suppress the resonant laser scatter without losing the single photons.

%%%%%%%%%%                       Figure 1                  %%%%%%%%%%%%%%%%%%%%
\begin{figure*}
\includegraphics{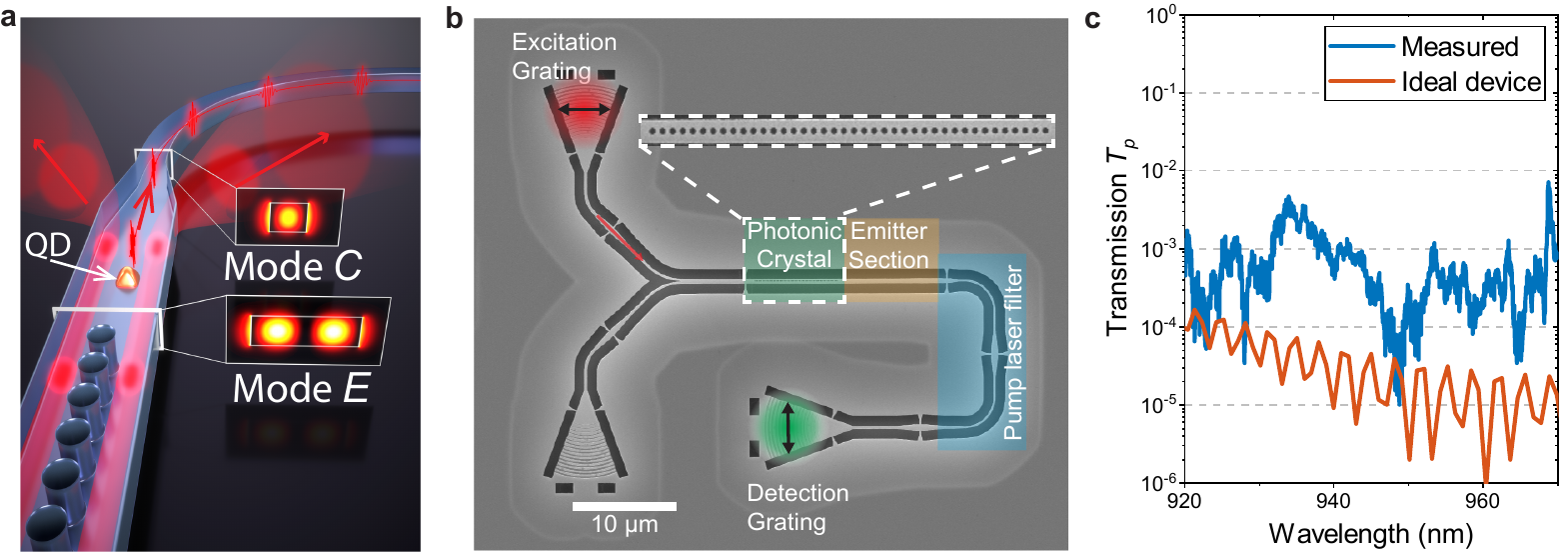}
\caption{\label{fig1}
Waveguide-based excitation scheme. (a) Illustration of the mode filtering operation. The resonant pump laser in the first-order waveguide mode excites the emitter and is subsequently squeezed out of the waveguide in the taper section. The QD emission into the fundamental mode of the waveguide is collected efficiently and guided. The photonic crystal acts a mirror for the fundamental mode, thereby enabling the directional out-coupling of the QD signal. (b) Scanning electron microscope image of the fabricated device. The excitation and collection spots are highlighted (red and green spots). The Y-splitter is used to excite the fundamental and first-order modes of the waveguide. The photonic crystal (zoomed in the inset to highlight the lattice of air holes) selectively transmits only the first-order mode into the emitter section. The pump laser filter section is composed of a waveguide taper and two 90$^\circ$ bends to suppress the pump laser. The bottom-left grating is used to align the in-coupling of the laser beam by monitoring the reflected signal from the photonic crystal. (c) The measured and calculated transmission $T_p$ spectrum of the device for a laser coupled in at the excitation grating and collected at the detection grating.}
\end{figure*}
%%%%%%%%%%%%%%%%%%%%%%%%%%%%%%%%%%%%%%%%%%%%%%

The operational principle of the device is presented in Fig. \ref{fig1}(a).
We design a two-mode nanophotonic waveguide where the embedded QD is efficiently coupled to the fundamental mode and weakly coupled to the first-order mode.
By selectively launching the laser into the first-order mode (excitation mode $E$), the QD is excited and the single-photon emission collected through the fundamental mode (collection mode $C$).
In order to efficiently collect only the single photons, the residual excitation in laser mode $E$ must be filtered out, while ensuring lossless propagation of mode $C$.
An adiabatically tapered waveguide section is employed to satisfy these demands simultaneously.
In the taper section, the $E$ mode becomes leaky and is extinguished by the deliberate introduction of sharp waveguide bends.
The adiabatic taper ensures the efficient transfer of the $C$ mode into the single-mode regime that subsequently can be coupled into an optical fiber.
We furthermore employ a one-dimensional photonic crystal as a backward reflector for single photons propagating in the mode $C$ to maximize unidirectional outcoupling efficiency.
A scanning electron microscope image of the nanofabricated device highlighting the three key elements of the device (photonic crystal, two-mode waveguide with emitters, and waveguide-taper-based pump laser filter) is shown in Fig.~\ref{fig1}(b) (see Sec.1 of Supplementary Information for the fabrication method).
Three high-efficiency ($>$ 65\%) grating couplers \cite{zhou2018} are fabricated for in- and out-coupling of light from free-space to the device.

The input excitation grating is connected to a 300-nm-wide single-mode waveguide, followed by a Y-splitter that launches the excitation laser into both the $E$ and $C$ modes of the two-mode waveguide \cite{syms_optical_1992}.
The Y-splitter together with the photonic crystal selectively prepares the mode of the excitation pulse (see Sec. II of Supplementary Information).
The photonic-crystal section is a key design element of the device serving a dual purpose: 1) as a backward reflector for unidirectional collection of single-photon emission and 2) to selectively prepare the excitation laser in the mode $E$.
It is designed such that it reflects the $C$ mode and transmits the $E$ mode into the emitter section of the waveguide.
Figure \ref{fig1}(c) shows the measured transmission spectrum $T_p(\lambda)$ of the excitation laser through the device, which quantifies how well the residual excitation light can be suppressed.
$T_p$ is extracted by comparing the transmitted laser intensity in two nominally similar devices with and without the photonic crystal section.
For reference, the calculated performance for an ideal device without any fabrication imperfections is shown in Fig. \ref{fig1}(c), and remarkably ideal performance with $T_p \sim 10^{-5}$ is observed in certain wavelength bands.
The minor deviations in the measurements from ideal performance can be attributed to an unintentional disorder in the nanofabricated photonic crystal.

%%%%%%%%%%                       Figure 2                  %%%%%%%%%%%%%%%%%%%%
\begin{figure}
\includegraphics{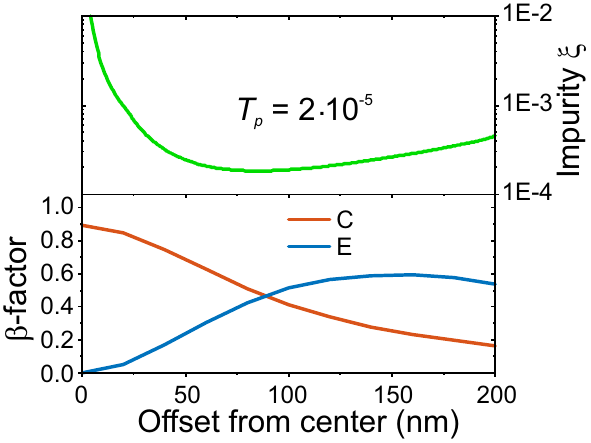}
\caption{\label{fig2} Predicted device performance. Top panel: expected single-photon impurity $\xi$ for the experimentally achieved value of $T_p$ = 2$\cdot$10$^{-5}$ and as a function of different emitter locations in the waveguide. Bottom panel shows the calculated $\beta$-factor for the two waveguide modes as a function of the offset distance of the emitter from the center of the waveguide. }
\end{figure}
%%%%%%%%%%%%%%%%%%%%%%%%%%%%%%%%%%%%%%%%%%%%%%%%

In order to assess the performance of the device as a single-photon source, the laser suppression $T_p$ should be related to the single photon emission probability.
An essential figure-of-merit is the intensity of the residual pump intensity relative to the intensity of the emitted single photon signal, i.e. the single-photon impurity $\xi$, which is the fraction of laser photons per single photons.
$\xi$  is related to the measured second-order coherence function through $g^{(2)}(\tau = 0) = 2\xi - \xi^2$ \cite{kako2006}.
We relate $T_p$ and $\xi$ as follows:
The residual laser intensity at the outcoupling grating is given by $I_p T_p$, where $I_{p}$ is the input pump laser intensity.
Under pulsed resonant excitation, we express the single-photon intensity at the collection grating as $I_\mathrm{sp} \approx \beta_E \beta_C I_p /2 $, which is a simplified expression for clarity that holds below saturation of the QD and when omitting any effect of dephasing.
The factor of $1/2$ accounts for the power splitting of the excitation laser into the modes $E$ and $C$ at the Y-splitter.
Section VI of the Supplementary Information details the complete theory without these restrictions.
$\beta_E$ and $\beta_C$ are the photon $\beta$-factors \cite{Arcari2014} expressing the probability of the QD to absorb a pump photon and emit a single photon into the waveguide, respectively.
Consequently we have
\begin{equation}
\xi = \frac{I_p T_p}{I_{sp}}=\frac{2T_p}{\beta_E\beta_C}.
\label{eq1}
\end{equation}
The QD position affects the emitter-photon coupling $\beta_C$ and $\beta_E$ and therefore the value of $\xi$.
Figure \ref{fig2} (bottom panel) shows the calculated $\beta$-factors as a function of transverse offset from the waveguide center.
A QD positioned exactly at the center of the waveguide maximally couples to $\beta_C$, but is not pumped by the excitation laser in the mode $E$ as $\beta_E \sim 0$.
The optimum QD position that simultaneously minimizes the photon impurity $\xi$ and maintains a high $\beta_C$ is seen in Fig.~\ref{fig2} for the measured device parameters, ($T_p$ = 2 $\cdot$ 10$^{-5}$).
For a QD position where a high single-photon coupling efficiency of $\beta_C\simeq0.9$ can be reached, we obtain $\xi\simeq5\cdot10^{-4}$, which implies that $g^{(2)}(0)\simeq 10^{-3}$ can be achieved.
We note that further reduction in $T_p$, e.g., by optimizing the filter design, could lead to even better single-photon purity even when $\beta_C$ approaches unity.

%%%%%%%%%%                       Figure 3                  %%%%%%%%%%%%%%%%%%%%
\begin{figure*}
\includegraphics{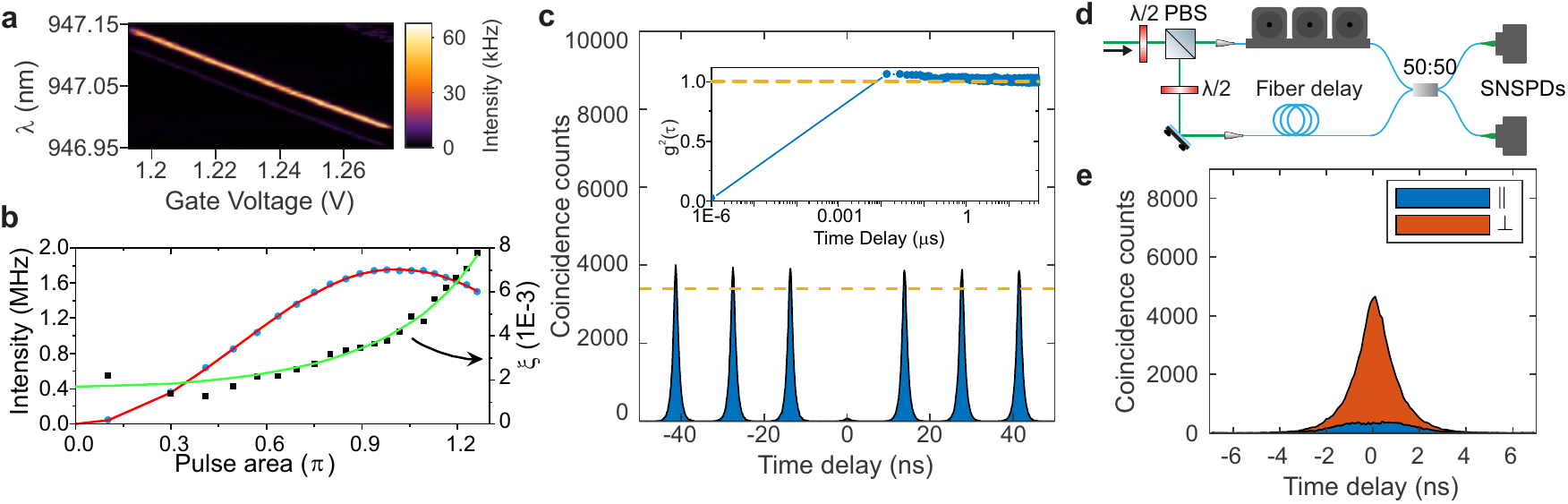}
\caption{\label{fig3} Demonstration of pure and indistinguishable single photons by pulsed deterministic resonant excitation.  (a) QD Resonance fluorescence intensity under cw laser excitation at a power of $P = 0.01\cdot P_{sat}$, where $P_{sat}$ is the saturation power. (b) Power dependence of the resonance fluorescence intensity andthe photon impurity $\xi$. The Rabi-oscillations (red curve) of the two-level system are modeled including a pure dephasing rate of $\gamma_{d} = 0.2$ ns$^{-1}$. (c) The intensity-correlation histogram in a Hanbury Brown and Twiss experiment for $\pi-$pulse excitation. The second order correlation function $g^{(2)}(0) = 0.02 \pm 0.005$ is extracted from the fitted amplitude of the central peak relative to the fitted amplitude for peaks at a time delay of $50$ $\mu$s (dashed line). The inset shows $g^{(2)}(\tau)$ measured by integrating the coincidences under the peak over the 50 $\mu$s timespan. (d) Schematic of the Hong-Ou-Mandel interferometer used for measuring the indistinguishability of two subsequent photons delayed by the laser pulse separation of 13.7 ns. (e) Coincidence counts after the Hong-Ou-Mandel interferometer when the input photons are co-polarized (blue) and cross-polarized (red).}
\end{figure*}
%%%%%%%%%%%%%%%%%%%%%%%%%%%%%%%%%%%%%%%%%%%%%%%%

Experimental demonstration of waveguide-assisted pulsed resonant excitation of an optimally coupled QD was demonstrated on the device shown in Fig. \ref{fig1}(b).
Resonance-fluorescence measurement from a neutral exciton under continuous wave excitation is shown in Fig. \ref{fig3}(a), which is carried out to identify QD resonances and demonstrate low-noise performance.
We observe distinct QD resonances, free of excitation laser background ($T_p$ = 2$\cdot$10$^{-5}$), with a linewidth of 800 MHz that tune with the applied bias voltage.
The broadening of the QD resonances beyond the natural linewidth (250 MHz, as estimated from lifetime measurement) occurs primarily due to slow spectral diffusion (time scale of 10 ms), which is not relevant for pulsed operation and could be rectified by active feedback \cite{Prechtel2013,Atature2014}.
Deterministic pulsed resonant excitation is performed with 26 ps optical pulses tuned to the QD resonance.
The observed Rabi oscillations of the detected intensity are shown in Fig. \ref{fig3}(b) that are modelled as a driven two-level system including minor pure dephasing, see Sec. VI of the Supplementary Information for details of the model.
The single-photon impurity $\xi$ was extracted at each excitation power by comparing the detected intensity with the QD tuned on- and off-resonance by using the electrical control.
The power-dependent $\xi$ reflects the fact that the QD transition saturates when approaching $\pi$-pulse excitation while the residual laser background scales linearly with pump power, and this behavior is fully captured by the theoretical model, cf. Fig. \ref{fig3}(b).
The coupling efficiency of the QD emission to the waveguide, quantified through $\beta_C$, is extracted by comparing the measured $\xi$($P\rightarrow0$) = 1.7$\cdot$10$^{-3}$ and $T_p$ = 2$\cdot$10$^{-5}$ values with the calculations in Fig. \ref{fig2}.
This comparison results in $\beta_C$ = 0.8, which corresponds to a QD position offset from center of the waveguide by $\approx$ 20 nm.
Hence, the device enables 80\% collection efficiency of the single photons into the waveguide while ensuring low laser background.
At $\pi$-pulse, i.e. deterministic QD preparation, we find $\xi$ = 0.004 ($g^{(2)}$($0$) = 0.008).
We detect a single-photon rate of 1.8 MHz, which is primarily limited by the collection optics in the device characterization setup and can readily be improved further. Section III of the Supplementary Information presents a detailed description of the observed source efficiency that fully accounts for the independently measured parameters.

Having demonstrated pulsed resonant excitation through the waveguide mode, we proceed to the characterization of the quality of the single-photon source.
Figure \ref{fig3}(c) shows the intensity-correlation histogram measured at $\pi$-pulse excitation using a Hanbury Brown and Twiss interferometer.
A clearly suppressed peak at time delay $\tau = 0$ ns is observed that is normalized to the long $\tau$ limit to extract $g^{(2)}(\tau = 0) = 0.020 \pm 0.005$.
The observed value of $g^{(2)}(0)$ is higher than the expected value for the measured device parameters, which can be attributed to the temporal extent of the excitation laser pulses (26 ps) in comparison to the QD decay time (640 ps) that results in non-zero two-photon emission probability\cite{fischer2017}.
We estimate that excitation laser with $<$ 3 ps pulse width would be required to reach the $g^{(2)}(0)$ value limited by the device \cite{fischer2017,sumanta2019}.
Even better peformance could be achieved by reducing $T_p$ either by design or through an improvement in the fabrication.
The current design enables  $T_p \approx 10^{-6}$ (see Fig. \ref{fig1}(c)) corresponding to $g^{(2)}(0) \approx 10^{-4}$, which approaches the best reported value in the literature obtained with two-photon resonant excitation \cite{zwiller2018}, where pump filtering is not a challenge.

Most applications of single photons in quantum information require high indistinguishability of the photons, which we measure in a Hong-Ou-Mandel (HOM) experiment by interfering two subsequently emitted photons in an unbalanced fiber-based Mach-Zehnder interferometer, cf. Fig. \ref{fig3}(d).
Figure \ref{fig3}(e) shows the recorded correlation histogram between the two detectors, where the strong suppression of coincidences for zero detector time delay testifies the high degree of indistinguishability of the emitted photons.
By controlling the polarization of the incoming photons the reference case of fully distinguishable photons (perpendicular polarization case) is recorded and we extract the HOM interference visibility $V$ that quantifies the photon indistinguishability.
We measure a raw visibility of $V_\textrm{raw} = (91 \pm 2)\%$, which, after correcting for the finite $g^{(2)}(0)$ and setup imperfections corresponds to $V = (96 \pm 2) \%$ (see Sec. IV of the Supplementary Infomation for details).
The measured indistinguishability is on par with the best reported value with cross-polarized resonant excitation \cite{Wang2016} and only superseded by experiments relying on excitation pulse-engineering \cite{Wei2014,pan2019dichromatic}.

In conclusion, we have experimentally demonstrated an efficient waveguide circuit for deterministic pulsed resonant excitation of QDs embedded in planar photonic nanostructures.
The circuit enables the realization of an efficient `plug-and-play' single-photon source featuring near-unity single-photon coupling, as well as high purity and indistinguishability.
The robust excitation process implies that the device could be operated continuously without any realignment, and as a proof-of-concept we operated the source hands-free for over $10$ hours with less than $2\%$ fluctuation in the generation rate (see Sec. V of the Supplementary Information).
The device will also enable improving the collection efficiency for more advanced excitation schemes relying on dichromatic laser pulses \cite{pan2019dichromatic}, which are typically limited by low-efficiency spectral filters.
An obvious next step is to implement direct chip-to-fiber coupling \cite{Ugurlu2019} thereby circumventing loss associated with collection, mode shaping and subsequent fiber coupling.
Another opportunity is to scale-up the circuit so that one excitation pulse could be pumping multiple QDs in parallel.
With such an approach the benefits of the scalable planar platform will be fully exploited in the ongoing pursuit of scaling up single-photon technology\cite{pan2019boson}.

\appendix
\section{Heterostructure composition and sample fabrication}
\begin{figure}
\begin{center}
\includegraphics{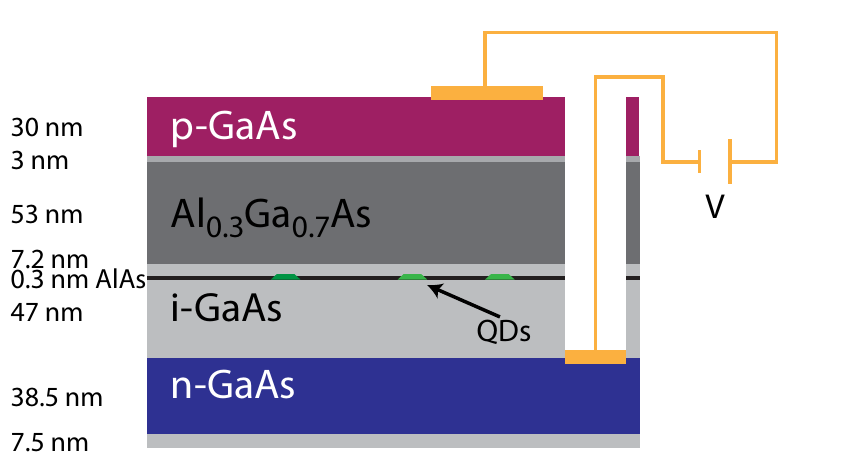}
\caption{\label{figS1} Outline of $p$-$i$-$n$ diode heterostructure used to realize the device.}
\end{center}
\end{figure}

The samples are fabricated on a GaAs membrane grown by molecular beam epitaxy on a (100) GaAs substrate. A 1150-nm-thick Al$_{0.75}$Ga$_{0.25}$As sacrificial layer is used to isolate and suspend the membrane from the substrate after wet etching. 
The membrane structure is shown schematically in Fig. \ref{figS1}. 
It contains a layer of self-assembled InAs quantum dots (QDs) grown with a technique that removes the electron wetting layer states \cite{Lobl2019}, embedded in a $p$-$i$-$n$ diode for the reduction of charge noise and control of the charge state and Stark tuning of the emitter.
A 53-nm-thick Al$_{0.3}$Ga$_{0.7}$As layer is used as a barrier to limit the current to a few nA when the diode is operated under forward bias.

Electrical contacts to the $p$-doped and $n$-doped layers are fabricated first. 
Reactive-ion etching (RIE) is used to open vias to the buried n-layer and Ni/Ge/Au contacts are deposited by electron-beam physical vapor deposition. 
The contacts are annealed at 430 $^\circ$C. Subsequently Cr/Au pads are deposited on the surface to form Ohmic $p$-type contacts. 
The waveguides are patterned using electron-beam lithography at 125 keV (Elionix F-125) and etched in the GaAs membrane by inductively-coupled plasma RIE in a BCl$_3$/Cl$_2$/Ar chemistry. 
The sample is then undercut and cleaned following the procedure explained in Ref. \cite{midolo_soft_2015}.

\section{Design of the photonic crystal section}
\begin{figure*}
\begin{center}
\includegraphics[height=7cm]{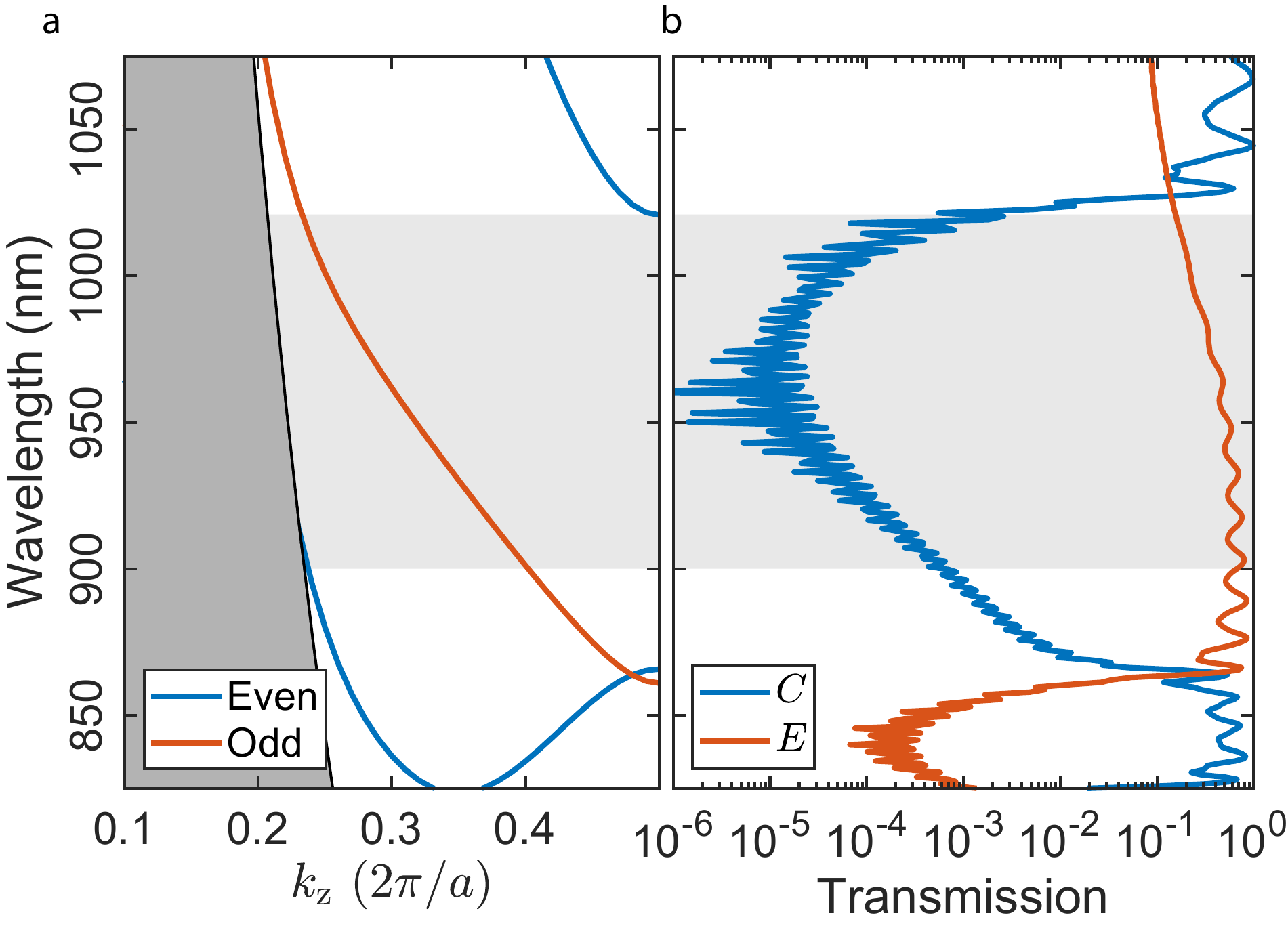}
\caption{\label{figS2} (a) Photonic bandstructure calculated for the multimode photonic crystal waveguide shows the behavior of the first odd and even modes allowed in the waveguide. The odd mode is labeled as $E$ as its employed to excite the quantum dot and the even mode is label $C$ as its employed to collect the QD emission. The dark gray shaded region is above the light line of the 170 nm thin slab. (b) Calculated transmission of the $E$ and the $C$ modes across a 20-hole photonic crystal waveguide section of 450 nm width, 170 nm thickness, and hole-to-hole distance (lattice constant) of 210 nm. The light gray shaded region in both the figures corresponds to the stop gap of an infinite photonic crystal.}
\end{center}
\end{figure*}
We employ a multimode photonic crystal nanobeam waveguide in the device to prepare the excitation laser in mode $E$ and as backward reflector for single photons in mode $C$.
The band structure for the multimode photonic crystal nanobeam waveguide (width = 450 nm) is shown in Fig.~\ref{figS2}(a) where $k_z$ is the projected wave-vector for propagation along the nanobeam waveguide.
The one-dimensional photonic crystal is realized as an array of circular air holes with a radius of 70 nm and a hole-to-hole spacing of 210 nm.
The solid curves below the light line (dark gray area) indicate the propagating modes confined in the nanobeam waveguide and are color coded according to their transverse spatial symmetry.
In the wavelength region highlighted in light gray, the photonic crystal supports a stop gap for even modes while allowing partial transmission of odd symmetry modes.
In our design, we couple the resonant pump laser into the odd mode that excites the QDs (labeled as excitation mode $E$) and collect the single photon emission through the even mode (labeled as collection mode $C$).
Under ideal conditions (lossless and infinitely long photonic crystal), light coupled to mode $C$ is completely reflected, while that in mode $E$ would be transmitted.

We employ finite-difference time-domain (FDTD) calculations of a 20-hole photonic crystal nanobeam waveguide to investigate the performance of a finite-length device. The results from the calculations are shown in Fig. \ref{figS2}(b). The excitation mode $E$ is attenuated by roughly 50\% over the spectral band of interest for the QDs (i.e. between 920--960 nm). The mode $C$ is instead extinguished by more than a factor of $10^4$ near the center of the band gap. Imperfections introduced during fabrication can reduce the total mode suppression. For this reason, the fabricated device has been designed with 40 holes.

In the transmission of the laser, this high degree of suppression of the $C$ mode by the photonic crystal filter allows preparing the resonant excitation laser selectively in the $E$ mode. In the collection of QD emission, the photonic crystal acts as a perfect ($>$99.99\% reflectivity) mirror for single photons coupled to the $C$ mode, thereby enabling unidirectional collection.

\section{Experimental setup}
In order to perform single photon generation experiments, the sample is cooled to a temperature of 1.6 K in a closed-cycle cryostat with optical and electrical access.
The excitation laser and the QD emission are focused and collected at the respective grating outcouplers (see Fig. \ref{fig1}(b)) using a wide field-of-view microscope objective.
A 20:80 (reflection:transmission) beam splitter is used to separate the excitation and collection into separate optical paths, with the high-efficiency path used for collection.
The collected single-photon emission is coupled into a single mode optical fiber and sent through a spectral filter constituting of an etalon (linewidth = 3 GHz; free spectral range = 100 GHz).
The spectrally filtered single photon stream can be directed to either a compact fiber-based unbalanced Mach-Zehnder for measuring two-photon interference or directly to a super-conducting nanowire single-photon detector (SNSPD).
The gate voltage across the QD is tuned using a low-noise voltage source with an RMS noise $<$50 $\mu$V, which corresponds to $<$0.1$\Gamma$, where $\Gamma$ is the linewidth of the QD.

\section{Samples}
A scanning electron microscope image of the nanofabricated device with a footprint of 50$\times$45 $\mu$m$^2$ is shown in Fig.~\ref{fig1}(b).
The photonic crystal section is a one-dimensional lattice of 40 air holes with radius of 70 nm and lattice spacing of 210 nm.
The emitter section (450-nm-wide and 170-nm-thin suspended GaAs nanobeam waveguide) supports the two propagating modes $E$ and $C$.
Self-assembled indium arsenide (InAs) QDs, embedded in a $p$-$i$-$n$ diode (see Supplemetary Fig. S1 for details), are randomly located across the waveguide with an average density of 10 $/\mu$m$^2$.
This density is high enough to comfortably find 3 - 4 QDs within the best laser suppression windows in all 20 fabricated devices.
The suspended waveguide is electrically contacted (contacts not shown in the figure) to tune the QDs and to suppress noise leading to spectral drift.
The pump laser filter is a 5-$\mu$m-long linear taper, which gradually reduces the waveguide width from 450 nm to 200 nm.
Two consecutive 90$^\circ$ waveguide bends are introduced to further extinguish the weakly-guided $E$ mode.
Three shallow-etched grating couplers are fabricated for in- and out-coupling of light from free-space to the waveguides.
These gratings enable $>$65\% collection efficiency of light in the $C$ mode from the waveguide into a single-mode optical fiber \cite{zhou2018}.

\section{Source efficiency}
\begin{figure}
\begin{center}
\includegraphics[width = \columnwidth]{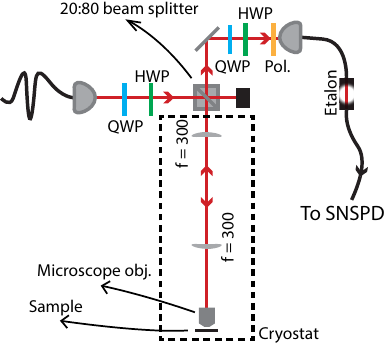}
\caption{\label{figS3} Schematic of the optical setup used in the excitation and collection of emission from a QD embedded in the nanophotonic device. The sample is cooled to a temperature of 1.6 K in a closed-cycle cryostat. A set of quarter (QWP) and half (HWP) wave plates are used to control the polarization of the incident and collected light.}
\end{center}
\end{figure}

The optical setup employed in our experiments is shown in Fig. \ref{figS3}.
The transmittance of each optical element used in the setup is carefully characterized using a continuous-wave narrow bandwidth diode laser operating at $947.1$ nm.
The complete breakdown of the source efficiency into the collection and QD efficiencies is presented in Table \ref{tablEff}.
A resonant excitation laser is collimated and imaged to the back focal plane of a low-temperature compatible microscope objective (NA = 0.81).
The microscope objective couples the laser light into the excitation grating coupler as well as collects the QD emission at the grating outcoupler.
The resonant laser and the collected emission is separated into different spatial modes using a 20:80 (reflection:transmission) beam splitter, where the transmission arm is used for collection.
The collected emission passes through a set of quarter and half wave plates (QWP, HWP in the figure) and is imaged onto a fibre collimator.
The collection efficiency of the imaging system $T$ from the device to the entrance of the collection fibre is $51\pm2\%$.
The QD emission coupled into the waveguide is fibre-coupled using the grating outcoupler-fibre relay.
In the current setup, the mode-matching efficiency of the grating outcoupler to the fibre $\eta_f$ is limited to $24\pm2 \%$, which is significantly lower than the $>65\%$ reported in our earlier work \cite{zhou2018}.
This is due to an unoptimized image relay line in the current setup that could readily be improved by a proper lens choice.

\begin{figure}
\begin{center}
\includegraphics[width=\columnwidth]{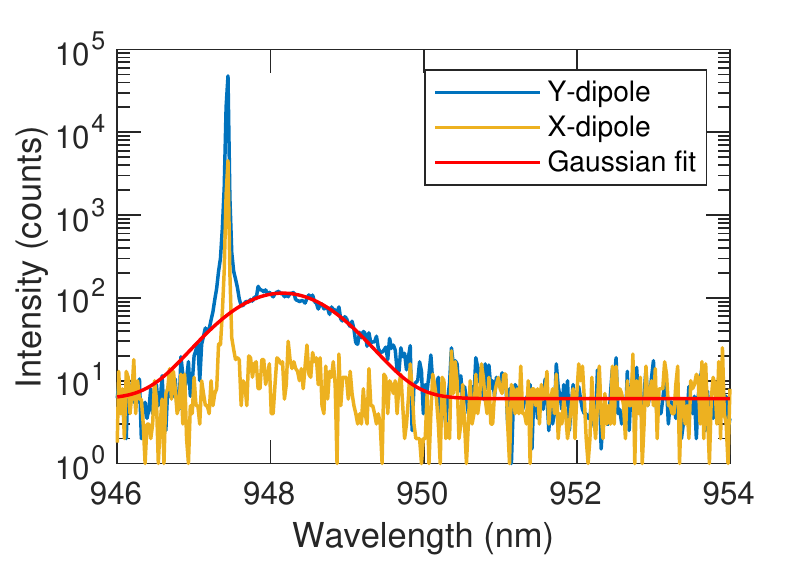}
\caption{\label{figS4} Spectrally-resolved resonance fluoresence of the QD excited using a narrow bandwidth diode laser. The emission spectra are collected at an excitation power of $1$ nW and gate voltage of $1.235$V with the laser tuned to $\lambda$ = 947.075 nm (X-dipole; yellow curve) or 947.108 nm (Y-dipole; blue curve). The red curve is the gaussian fit of the emission in the photon side band when exciting the $Y$-dipole.}
\end{center}
\end{figure}

\begin{table*}
\begin{center}
\begin{tabular}{|c|l|c|}
\hline
\multirow{4}{*}{QD efficiency}& Y-dipole fraction ($\eta_Y$) &  $91 \pm 1\%$\\
& Filtered  phonon sideband $(1-\eta_\textrm{ZPL})$ & $8.5 \pm 0.5\%$\\
& QD blinking $(1-\eta_\textrm{blink})$ & 3\%\\
& $\beta-$factor $(\beta_C)$ & $80 \pm 5\%$ \\

\hline
\multirow{4}{*}{Collection efficiency} & On-chip propagation loss $(1-\eta_p)$ & $15 \pm 5\%$\\
& Collection optics $(T)$ & $51 \pm 2 \%$\\
& Grating-to-fibre collection $(\eta_{f})$ & $24 \pm 2\%$\\
& Spectral filter $(\eta_s)$ & $80 \pm 1\%$  \\
\hline
Total efficiency & & $5.3 \pm 0.7 \%$\\
\hline
Detection efficiency &  & $65 \pm 5\%$\\
\hline
& Laser rep. rate & 72.5 MHz\\
& Expected rate & 2.5 $\pm$ 0.4 MHz\\
& Detected rate & 2.2 MHz\\
\hline
\end{tabular}
\end{center}
\caption{This table presents the end-to-end efficiency of the single photon source.}
\label{tablEff}
\end{table*}

In the following, we analyze the emission efficiency of the QD employed in the measurements taking fully into account all relevant loss processes.
We operate the QD at a gate voltage of 1.235 V, which ensures selective excitation of the neutral exciton $X_0$.
$X_0$ has two bright states from spectrally non-degenerate dipoles (fine structure splitting = 10 GHz) with orthogonal linear polarization.
The transverse location of QD in the waveguide determines the coupling of the dipoles to the waveguide modes.
The coupling asymmetry of the two dipoles is estimated from the ratio of the resonance fluorescence intensities at a fixed excitation power of 1 nW ($ \approx 1\%$ of the saturation power for the well-coupled $Y$-dipole).
The spectrally resolved emission with the excitation laser on resonance with $X$- and $Y$-dipoles is shown in Fig. \ref{figS4}.
Under pulsed resonant excitation, both the dipoles are driven by the broad spectral bandwidth (40 GHz) of the pulse, and the coupling asymmetry relates to the emission asymmetry in the two dipoles.
By integrating the area under the spectrum, we estimate that the emission fraction into the well-coupled $Y$-dipole is $\eta_Y$ = 91 $\pm$ 1\%.
The resonance fluorescence spectrum in Fig. \ref{figS4} also exhibits a weak pedestal, which corresponds to the residual phonon side band.
The phonon sideband is fitted to a gaussian to estimate the fraction emitted outside the zero phonon line, $1-\eta_\textrm{ZPL}$ = 8.5 $\pm$ 1\%.
Apart from these radiative losses outside the zero phonon line of the $Y$-dipole, the QD neutral exciton can weakly couple to non-radiative dark state.
This contribution is obtained by modeling the weak bunching observed (cf. data in Fig. 3(c) in the main text) of maximum amplitude $\max(g^{(2)}(\tau))/g^{(2)}(\tau \rightarrow \infty) = 1.03$ with an exponential decay rate of $0.25 \mu s^{-1}$ using a 3-level system and extracting the dark state population \cite{Johansen2010,Davanco2014}.
The resulting probability for the QD to blink into the dark state is $1-\eta_\textrm{blink}$ = 3\%.
The product of $\eta_Y$, $\eta_\textrm{ZPL}$, and $\eta_\textrm{blink}$ is the intrinsic efficiency of the QD, which is 80 $\pm$ 1\%.
The coupling efficiency of the QD emission in $Y$-dipole to the waveguide collection mode $C$, quantified through $\beta_C$, is found to be 80 $\pm$ 5\%, as discussed in the main text.
The collected emission is spectrally filtered using an etalon filter with a linewidth of $3$ GHz (peak transmission efficiency $\eta_s$ = 80 $\pm$ 1\%) centered at the $Y$-dipole emission wavelength so as to filter out the phonon side band and the X-dipole.
The total end-to-end efficiency of the source is $\eta_Y \eta_\textrm{ZPL}\eta_\textrm{blink}\beta_C\eta_p T \eta_f\eta_s$, where all the measured contributions to propagation loss from source to detector are listed in Table \ref{tablEff}. 
Specifically the minor residual loss inside the device due to propagation in the waveguides was measured by dedicated transmission measurements through waveguides of varying lengths. 
The estimated propagation loss in the waveguide is $10.5$ dB/mm, which for the $\approx 100\mu$m long device results in a loss (1-$\eta_p$) of $\approx 15 \pm 5\%$.
The overall efficiency of the source was found to be $5.3\%$, and the complete break-down of the efficiency lays out straightforward path ways to improve this further. 
The detected and expected photon count rates listed in the table take into account the detector deadtime of 100 ns. 
Notably the detected rate of single photons match the expected rate to within the error bars of the measured parameters, emphasizing the full quantitative understanding of the device.

\section{Analyzing indistinguishability data and setup parameters}
We employ the procedure discussed in Ref.~\onlinecite{Liu2018} and correct the raw indistinguishability for setup imperfections and finite $g^{(2)}(0)$.
The raw coincidence counts, shown in Fig. 3(e) of the main text, are fitted with double sided exponentials convoluted with the measured instrument response function of the detectors.
To account for the low background count, we employ a Poissonian noise model with amplitudes of the central peak $A_0$ and the peak at long time delay $A_{\infty} $\cite{Kirsanske2017}.
The fitted peak amplitude at zero-delay $A_0$ is rescaled to $A_\infty$. This procedure is carried out to extract the normalized central peaks $A_\perp$ and $A_\parallel$ for co- and cross-polarized photons, respectively. The normalization procedure corrects for systematic variations in the total count rates that could occur when switching between the two configurations. The normalized areas are related to the raw indistinguishability $V_\mathrm{raw}$ as
\begin{equation}
	V_\mathrm{raw} = \frac{A_\perp-A_\parallel}{A_\perp}\label{eq:Vraw}.
\end{equation}

\begin{figure}
\begin{center}
\includegraphics{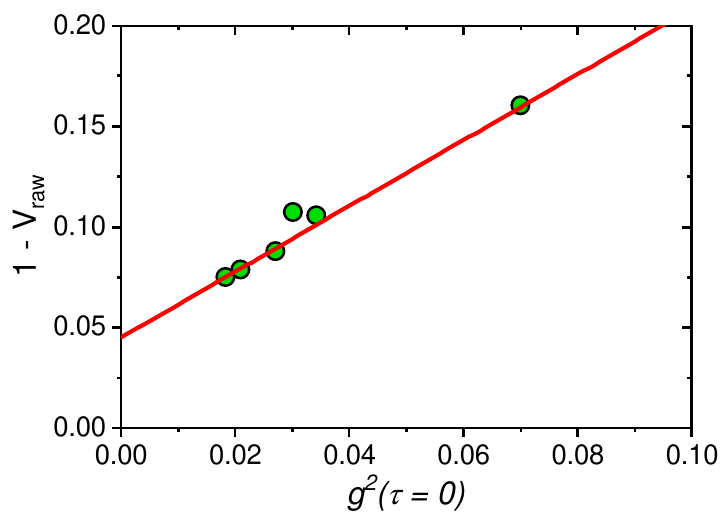}
\caption{\label{figS5} Relation betwen the QD signal to laser extinction for extracting the photon indistinguishability. At a constant excitation power, laser extinction was tuned by increasing the background scatter at the collection fibre. The plot shows measurements of $g^{(2)}(0)$ and the HOM visibility $V_\textrm{raw}$ at a given laser extinction (circles). The y-intercept of the linear fit (red curve) to the data is used to estimate the intrinsic HOM visibility, which is found to be $V = 96 \pm 2\%$.}
\end{center}
\end{figure}

For an intrinsic indistinguishability of $V$ the expected amplitude of the central peak accounting for setup imperfections is given by
\begin{equation}
A(V) = \left(R^3 T+R T^3\right)[1 + (2-\eta_\textrm{opt})g^{(2)}(0)]-2R^2T^2(1-\epsilon)^2 V,\label{eq:Vth}
\end{equation}
where $R$ and $T$ is reflectivity and transmission of the beam splitter, $(1-\epsilon)$ is the classical visibility of the interferometer, and $\eta_\textrm{opt}$ is the total optical efficiency of the setup shown in Tab. \ref{tablEff}. Using Eq.~\eqref{eq:Vraw} and \eqref{eq:Vth}, we can estimate the instrinsic visibility $V$ from $V_\mathrm{raw}$ using the relation
\begin{equation}
V = \frac{[1+(2-\eta_\textrm{opt})g^{(2)}(0)]\left(R^2+T^2\right) V_\mathrm{raw}}{2RT(1-\epsilon)^2}.
\label{eq:Vfull}
\end{equation}

In our experiment, we measured $R=0.476$, $T=0.524$, and $g^{(2)}(0) = 0.02\pm 0.005$, $V_\mathrm{raw}=0.91\pm0.02$, and $(1-\epsilon)>0.95$. These values for the setup parameters results in an intrinsic visbility of $V = 0.97 \pm 0.03$.

We also employed an alternative approach to correct the HOM visibility. In the setup, we employ a quarter-wave plate and a linear polarizer in the collection path to optimally collect the light at the grating outcoupler.
This polarization configuration also helps in suppressing the residual laser scatter that does not couple to the waveguide at the input grating.
By varying the position of the quarter-wave plate, we can vary the laser background in the setup.
At each position of the waveplate, we measure the $g^{(2)}(0)$ and $V_\textrm{raw}$. Fig. \ref{figS5} shows the measured raw HOM visibility (plotted as $1-V$) plotted against the measured $g^{(2)}(0)$.
We fit the data to a first-order polynomial following Eq. \ref{eq:Vfull}.
The y-intercept is the intrinsic HOM visibility of the source.
Using this approach, we estimate $V = 0.96 \pm 0.02$.

\section{Long-term operational stability of the plug-and-play source}
\begin{figure}
\begin{center}
\includegraphics[width = \columnwidth]{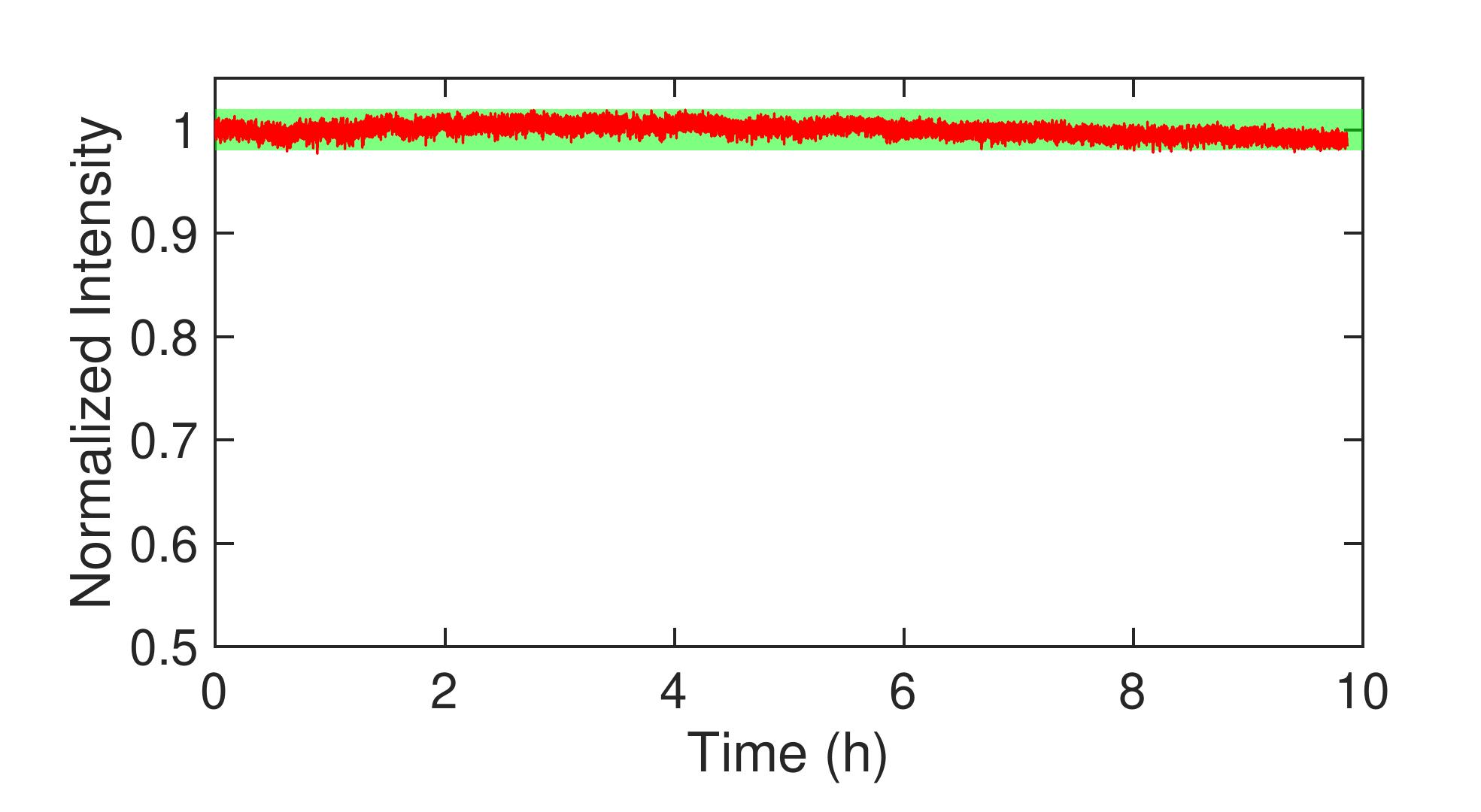}
\caption{\label{figS6} Detected single photon rate from the device over 10 hours of unmanned operation. We observed similar stability over several days, but without acquiring continuous measurements. The detected photon rate fluctuations are $<2\%$ (green shaded area), most of which are due to the slow polarization variations in the single mode fibre relay to the superconducting nanowire single-photon detector (SNSPD).}
\end{center}
\end{figure}

The long-term operational stability of the single photon source is monitored by continuously measuring the generated photon rate over 10 hours without any realignment of the setup.
The detected count rate is shown in Fig. \ref{figS6}, which highlights that the source exhibits $<2\%$ fluctuations in intensity over the whole measurement interval.
The residual slow variations in the count rates are primarily due to the long timescale thermal drifts in the single mode optical fibre that rotates the polarization of the single photons detected at the SNSPD.
The SNSPD can exhibts upto $12\%$ change in detection efficiency if the linear polarization of the single photons is changed from horizontal to vertical.
Hence, we attribute the long term drifts in the count rates to the polarization drift, rather than the stability in fibre outcoupling from the device.
We performed several such 10 hour acquisitions over 7 days with no obvious reduction in the emitted single photon rate.

\section{Theoretical model for resonance fluorescence}
We model the QD as a two-level system with ground state $\ket{g}$ and an excited state $\ket{e}$, with the frequency difference $\omega_\textrm{qd}$.
Defining the atomic raising and lowering operators $\hat{\sigma}_+ = \ket{e}\bra{g}$ and $\hat{\sigma}_- = \ket{g}\bra{e}$, respectively, we can write the non-interacting two-level system Hamiltonian as $\hat{H}_\textrm{qd} = \hbar \omega_\textrm{qd} \hat{\sigma}_+ \hat{\sigma}_-$.
We follow the derivation in \cite{Muller2007} to calculate the resonance fluoresence signal from the QD.
By driving the QD using a monochromatic field $\mathcal{E} = \mathcal{E}_0 e^{-i \omega_p t}$, where $\omega_p$ is the laser frequency that may be detuned from the QD by $\Delta = \omega_p - \omega_\textrm{qd}$, we can write the equation of motion for the resonantly excited QD with a radiative decay rate $\gamma$ and dephasing rate $\gamma_\textrm{d}$ as
\begin{equation}
\dot{\mathbf{\rho}}(t) = \mathbf{M} \cdot \mathbf{\rho}(t),
\label{eq:motion}
\end{equation}
where
\begin{equation}
\mathbf{M} = \begin{pmatrix}
0 & i\Omega/2 & -i\Omega/2 & \gamma \\
i\Omega/2 & -\frac{\gamma + 2\gamma_d}{2} + i \Delta & 0 & -i\Omega/2 \\
-i\Omega/2 & 0 & -\frac{\gamma + 2\gamma_d}{2} - i \Delta & i\Omega/2 \\
0 & -i\Omega/2 & i\Omega/2 & -\gamma \\
\end{pmatrix}
\end{equation}
and
\begin{equation}
\mathbf{\rho}(t) = \begin{pmatrix}
\rho_\textrm{gg}(t)\\
\rho_\textrm{ge}(t)\\
\rho_\textrm{eg}(t)\\
\rho_\textrm{ee}(t)\\
\end{pmatrix}.
\end{equation}
Here, $\Omega$ is the Rabi frequency, $\rho_\textrm{gg}(t)$ and $\rho_\textrm{ee}(t)$ are the ground and the excited state populations, respectively, and $\rho_{ge}(t)$ and $\rho_{eg}(t)$ denotes the coherence between the states.

Under pulsed resonant excitation of a QD with a gaussian pulse, the Rabi frequency can be represented as
\begin{equation}
\Omega(t) = \frac{\Theta}{\sqrt{\pi}\sigma} e^{-(t-t_0)^2/\sigma^2},
\end{equation}
where, $\Theta$ is the pulse area related to the excitation intensity as $\Theta \propto \sqrt{I_p}$ and $\sigma$ is the half-width of the pulse. The time offset, $t_0$, is the center of the pulse.

We solve the system of differential equations (Eq. \ref{eq:motion}) numerically for pulsed resonant excitation.

\subsection{Derivation of the single photon impurity $\xi$}
\begin{figure}
\begin{center}
\includegraphics[width = \columnwidth]{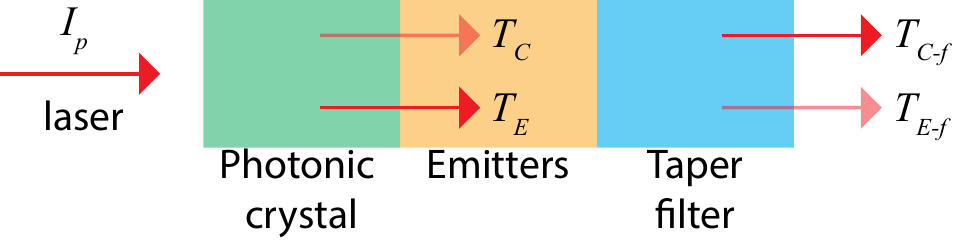}
\caption{\label{figS7} Operational schematic of the nanophotonic structure highlighting the three essential sections: the photonic crystal, emitter, and the taper filter. The propagation of the excitation laser across the various sections in the two waveguide modes $E$ and $C$ are highlighted. The opacity of the arrows indicates the transmission of the modes; brighter arrow for a higher transmittance value.}
\end{center}
\end{figure}

Figure \ref{figS7} illustrates the transport of the excitation laser with intensity $I_p$ through the device.
At the entrance of the photonic crystal section, the laser is equally coupled to the modes $E$ (excitation) and $C$ (collection) due to the Y-splitter design employed in the device shown in Fig. 1(b) of the main text.
The photonic crystal filter transmits a fraction $T_E$ of the mode $E$ and $T_C$ of the mode $C$.
Typical values for $T_E$ and $T_C$  are shown in Fig. \ref{figS2}.
In the emitter section, the excitation laser propagates unperturbed up to the entrance of the taper filter section.
The taper filter extinguishes the mode $E$ while transmitting the mode $C$.
The transmission of the modes $E$ and $C$ in the taper filter section are denoted $T_{E-f}$ and $T_{C-f}$, respectively.

The residual laser intensity $I_r$ at the output is expressed as
\begin{equation}
I_r = \frac{I_p}{2} \left(T_E T_{E-f} + T_C T_{C-f}\right) \equiv I_p T_p.
\end{equation}
The taper filters were numerically optimized to achieve $T_{E-f} \approx 10^{-6} - 10^{-7}$ and $T_{C-f} \approx 1$.
From Fig. \ref{figS2}(b) shows that the photonic crystal suppresses the transmission of exctation laser in the mode $C$ with $T_C \approx 10^{-5} - 10^{-6}$ at the operation wavelength of the device.

Under weak excitation of the QD, we can assume that the emitted single photon signal intensity $I_\textrm{sp}$ is proportional to the excitation intensity $I_p$.
We can express $I_\textrm{sp}$ as
\begin{equation}
I_\textrm{sp} = \frac{I_p}{2} T_E \beta_E \left( \beta_C + \beta_E T_{E-f} \right) + \frac{I_p}{2} T_C \beta_C \left( \beta_C + \beta_E T_{E-f} \right),
\label{sps}
\end{equation}
where we assumed $T_{C-f} = 1$. We can drop the last three terms as $T_{E-f},T_C \ll 1$ and express the collected single photon intensity as
\begin{equation}
I_\textrm{sp} = \frac{I_p}{2}T_E \beta_C \beta_E,
\end{equation}
which is the expression that we discuss in the main text. The impurity $\xi$ can then be expressed by assuming $T_E = 1$ as
\begin{equation}
\xi = \frac{I_r}{I_\textrm{sp}} \approx \frac{2 T_p}{\beta_E \beta_C}
\end{equation}

Using the numerical solution of Eq. \ref{eq:motion} to extract the relation between the single photon emission and $I_p$ in Eq. \ref{sps}, we can derive an exact expression for $I_\textrm{sp}$ and $\xi$, which is shown in Fig. 3(b) of the main text.

\begin{acknowledgments}
We gratefully acknowledge financial support from Danmarks Grundforskningsfond (DNRF) (Center for Hybrid Quantum Networks (Hy-Q)), H2020 European Research Council (ERC) (SCALE), Styrelsen for Forskning og Innovation (FI) (5072-00016B QUANTECH), Bundesministerium f\"{u}r Bildung und Forschung (BMBF) (16KIS0867, Q.Link.X), Deutsche Forschungsgemeinschaft (DFG) (TRR 160).
\end{acknowledgments}

%% Bibliography
%
\end{document}